\documentclass[12pt]{amsart}
\usepackage{amssymb,latexsym}
\theoremstyle{plain}
\numberwithin{equation}{section}
\newtheorem{theorem}{Theorem}

\begin{document}

\title[Minimal Operation Time of Energy Devices] {Minimal Operation Time of Energy Devices}
\author{Andreas Boukas}
\address{Department of Mathematics and Natural Sciences, American College of Greece\\
 Aghia Paraskevi, Athens 15342, Greece}
\email{andreasboukas@acgmail.gr}


\maketitle

\begin{abstract} \noindent We consider the problem of determining the minimal time for which  an energy supply source should operate in order to supply  a system with a desired amount of energy in finite time. 
\end{abstract}

\section{Introduction}

\noindent While boiling water or any other liquid, most of us have noticed that the heater can be switched off at an intuitively chosen time and the liquid will still reach its boiling point no long after the heater is switched off. A natural question arises: how can switch-off time be chosen in an optimal way so that electrical energy will not be wasted? In other words what is the earliest time at which the heater can be turned off while still reaching the liquid's boiling point in a finite time? A general formulation of the problem is as follows: Let D be a device that supplies energy to a system S through a supply line. Let $E^{\prime}(t)$ be the energy supply rate. We assume that D can be switched on and off and that it continues  to supply energy, at a decreasing rate, for some time after it has been switched off.    Question: What is the minimum switch-off time of D (corresponding to the minimal operational time of D) in order to transfer to S a total amount of energy $Q$ (where $Q>0$ is given)?  The device $D$ can also be viewed as a control mechanism for bringing the system $S$ from an energy level $E_1$ to a higher energy level $E_2$ in finite time while operating for the minimum time possible. Clearly, the solution of this problem can have a lot of applications , both civilian (e.g energy conservation) and military (e.g minimizing detection risk).

\section{Examples}

\noindent EXAMPLE 1. Exponential Model.

\smallskip

\noindent We consider a simple example of an energy device supplying  energy to a system at  time $t\geq0$ at a rate (which in what follows we consider to include the rate at which the transferred energy is possibly radiating from the system and/or the supply line)   given by

\[
E^{\prime}(t)=
\left\{
\begin{array}{ll}
e^{a\,t}-1&\mbox{if $0 \leq t \leq t_0$}\\
e^{a\,t_0}-1&\mbox{if $t_0 \leq t \leq t_1$}\\
\frac{e^{a\,t_0}-1}{1-e^{-b\,T}}\,\left(e^{-b\,(t-t_1)}-e^{-b\,T}\right)&\mbox{if $t \geq t_1$}
\end{array}
\right.
\]

\noindent where $a$ and $b$ are positive real numbers characteristic of the source but also depending on the environment, $t_0$ is the time at which the energy supply rate is at its peak, and $t_1$ is the switch-off time. We assume that the optimal switch-off time is after the rate of energy supply has been stabilized i.e that $\hat{ t_1}\geq t_0$.  We also assume that after switching-off at time $t_1\geq t_0$,  the source stops transferring energy to the device at time $t_1+T$,  where $T>0$ is independent of $t_1$.   Let $Q>0$ be the amount of energy that we wish to transfer to the system. We assume that the transfer of this amount of energy will occur at some time $t_2 \geq t_1\geq t_0$. The energy $Q$ could be, for example,  the energy required to bring a liquid substance to its boiling temperature,  or the energy required for complete phase transition. In the latter case $Q=m\,L_v$, where $m$ is the mass of the liquid and $L_v$ is its latent heat of vaporization (cf. [2]). We require that

\[ \int_0^{t_2}\,E^{\prime}(s)\,ds   =Q
\]

\noindent  which implies that 

\[\int_0^{t_0}\,E^{\prime}(s)\,ds +\int_{t_0}^{t_1}\,E^{\prime}(s)\,ds+ \int_{t_1}^{t_2}\,E^{\prime}(s)\,ds =Q
\]

\noindent or, by the definition of $E^{\prime}(s)$,

\[
\int_0^{t_0}\,  \left( e^{a\,s}-1 \right)  \,ds +\int_{t_0}^{t_1}\,      \left( e^{a\,t_0}-1 \right)        \,ds+ \int_{t_1}^{t_2}\,  \frac{e^{a\,t_0}-1}{1-e^{-b\,T}}   \,\left(e^{-b\,(s-t_1)} -e^{-b\,T}\right)      \,ds =Q
\]

\noindent which implies that

\[
 \left[ \frac{e^{a\,s}}{a}-s\right]^{s=t_0}_{s=0} + \left( e^{a\,t_0}-1 \right)\,(t_1-t_0)+   \frac{e^{a\,t_0}-1}{1-e^{-b\,T}}\,\left[-\frac{e^{-b\,(s-t_1)} }{b} -e^{-b\,T}\,s\right]_{s=t_1}^{s=t_2} =Q
\]

\noindent from which letting

\begin{eqnarray*}
y&:=&t_2-t_1\\
L_0&:=&\frac{1}{e^{a\,t_0}-1 }\,\left(Q+\frac{1}{a}-\frac{ e^{a\,t_0}}{a}+e^{a\,t_0}\,t_0+\frac{  1-e^{a\,t_0}}{1-e^{-b\,T}  }\,\frac{1}{b}\right)\\
L_1&:=&\frac{1}{b(1-e^{-b\,T})  }\\
L_2&:=&\frac{1}{1-e^{-b\,T}}
\end{eqnarray*}

\noindent we obtain

\begin{equation}
t_1= L_0+L_1\,e^{-b\,y}+L_2\,y.\label{ee1}
\end{equation}

\noindent Notice that  if $y=0$ then  $t_1=L_0+L_1$  is the time the device supplies the desired energy level $Q$ without being switched-off.   The maximum value of $y$ is  $y=y_{max}=(t_1+T)-t_1=T$. The optimal switch-off time $\hat{t_1}$  is therefore determined from (\ref{ee1}) by letting $y=T$ and it is given by 

\[
\hat{t_1}=L_0+L_1\,e^{-b\,T}+L_2\,T.
\]

\noindent Energy level $Q$ is reached at time

\[t_2=\hat{t_1}+T=L_0+L_1\,e^{-b\,T}+(L_2+1)\,T.\]

\smallskip

\noindent EXAMPLE 2. Linear Model.

\smallskip

\noindent A simplified version of the previous model is obtained by assuming that

\[
E^{\prime}(t)=
\left\{
\begin{array}{ll}
\frac{a}{t_0}\,t&\mbox{if $0 \leq t \leq t_0$}\\
a&\mbox{if $t_0 \leq t \leq t_1$}\\
-\frac{a}{T}\,(t-t_1)+a&\mbox{if $t \geq t_1$}
\end{array}
\right.
\]

\noindent where $t_0>0$, $a>0$, and $T$ is as in Example 1 above.  In this case

\[\int_0^{t_2}\,E^{\prime}(s)\,ds   =Q\]   

\noindent implies that

\[\int_0^{t_0}\,E^{\prime}(s)\,ds +\int_{t_0}^{t_1}\,E^{\prime}(s)\,ds+ \int_{t_1}^{t_2}\,E^{\prime}(s)\,ds =Q\]

\noindent or, by the definition of $E^{\prime}(s)$,

\[
\int_0^{t_0}\,\frac{a}{t_0}\,s \,ds +\int_{t_0}^{t_1}\, a\,ds+\int_{t_1}^{t_2}\, \left( -\frac{a}{T}\, (t-t_1)  +a  \right) \,ds =Q
\]

\noindent which implies that

\[
 \left[ \frac{  a}{t_0 }\,\frac{s^2}{2}\right]^{s=t_0}_{s=0} + a\,(t_1-t_0)+\left[-\frac{a}{2\,T}\,(s-t_1)^2+a\,s\right]_{s=t_1}^{s=t_2} =Q
\]

\noindent and letting  $y=t_2-t_1$ we obtain  

\[ 
t_1=\frac{1}{2\,T}\,y^2-y+\frac{Q}{a}+\frac{t_0}{2}.
\]

\noindent As in Example 1, substituting $y$ by  $ y_{max}=T$  we obtain

\[ 
\hat{t_1}=-\frac{1}{2}\,T+\frac{Q}{a}+\frac{t_0}{2}
\]
 
\noindent which is bigger or equal to $t_0$ if and only if  $\frac{Q}{a} -\frac{1}{2}\,T-\frac{t_0}{2}\geq0$.  Energy level $Q$ is reached at time

\[
t_2=\hat{t_1}+T=\frac{t_0}{2}+\frac{Q}{a} +\frac{1}{2}\,T.
\] 

\section{General Description of the Optimal Switch-Off Time}

\noindent The  examples treated in detail in the previous section suggest the following general theorems.

\smallskip

\begin{theorem}   Let, for each $t_1\geq t_0$

\[
E_{t_1}^{\prime}(t)=
\left\{
\begin{array}{ll}
f(t)&\mbox{if $0 \leq t \leq t_0$}\\
f(t_0)&\mbox{if $t_0 \leq t \leq t_1$}\\
g(t)&\mbox{if $t \geq t_1$}
\end{array}
\right.
\]

\noindent where $f$ is continuous and increasing with $f(0)=0$, and  $g$ is continuous and  decreasing with $g(t_1)=f(t_0)$. Let $F$ and $G$ denote the anti-derivatives of $f$ and $g$ respectively. If there exists $T>0$  such that $g(t_1+T)=0$ for all $t_1\geq t_0$,  then the optimal switch-off time $\hat{t_1}\geq t_0$  is  the solution of

\[F(t_0)+f(t_0)\,(\hat{t_1}-t_0)+G( \hat{t_1}+T  )-G(\hat{t_1})=Q.
\]

\end{theorem}

\begin{proof} The condition 

\[\int_0^{t_2}\,E_{t_1}^{\prime}(s)\,ds =Q\]

\noindent implies that 

\[\left(F(t_0)-F(0)\right)+f(t_0)\,(t_1-t_0)+\left(G(t_2)-G(t_1)\right)=Q\]

\noindent which by $F(0)=0$ and the fact that for the optimal $\hat{t_1}$,  $t_2=\hat{t_1}+T$ implies that 

\[F(t_0)+f(t_0)\,(\hat{t_1}-t_0)+G( \hat{t_1}+T  )-G(\hat{t_1})=Q.
\]

\end{proof}

\noindent We can generalize the above theorem to the case of a heat source that supplies energy at a possibly strictly increasing (or even arbitrary continuous) rate as follows.

\begin{theorem}    Let, for each $t_1\geq t_0$, the energy function  $E_{t_1}(t)$ be an increasing continuously differentiable function of $t$. The optimal switch-off time $\hat{t_1}$ and the associated time $t_2$ at which the energy level reaches $Q$ are determined by the system of equations

\[E_{\hat{t_1}}(t_2)=Q,\,\,\,\,\, E_{\hat{t_1}}^{\prime}(t_2)=0.
\]

\end{theorem}  

\begin{proof}  It is clear that the optimal switch-off time $\hat{t_1}$ makes full use of the energy source in the sense that the desired energy amount  $Q$ is supplied at the moment when the source dies out i.e when $E_{\hat{t_1}}^{\prime}(t_2)=0$. For such a $t_2$,

\[\int_0^{t_2}\,E_{  \hat{t_1}  }^{\prime}(s)\,ds =Q\]

\noindent implies

\[\int_0^{\hat{t_1}}\,E_{  \hat{t_1}  }^{\prime}(s)\,ds+ \int_{\hat{t_1} }^{t_2}\,E_{\hat{t_1} }^{\prime}(s)\,ds    =Q\]

 i.e $E_{  \hat{t_1}  }(  \hat{t_1})- E_{  \hat{t_1}  }( 0 )  +E_{\hat{t_1}  }( t_2)-E_{  \hat{t_1}  }(  \hat{t_1})    =Q$  and since $E_{\hat{t_1}}(0)=0  $, we obtain $E_{\hat{t_1}}(t_2)=Q$.

\end{proof}

\section{Noisy Supply Line}

\noindent The energy supply rate functions $\phi_{ t_1}(t):=E^{\prime}(t)$ used in Examples 1 and 2 above, can be viewed on each of the intervals $[0,t_0]$, $[t_0,t_1]$ and $[t_1,t_2]$,  as solutions of ordinary differential equations of the form

\begin{equation}
d\phi_{ t_1}(t)=(c_1\,\phi_{ t_1}(t)+c_2)\,dt\label{n1}
\end{equation}

\noindent for appropriate constants $c_1,c_2$ (depending on the interval) and initial condition $\phi_{ t_1}(0)=0$,  with the different branches of $\phi_{ t_1}(t)$ tied up at $t_0$ and $t_1$. It could be the case however, that the supply line connecting the energy source $D$ with the system $S$ is affected by noise, appearing in (\ref{n1}) in the form of additive noise 

\begin{equation}
d\phi_{ t_1}(t)=(c_1\,\phi_{ t_1}(t)+c_2)\,dt+ (c_3\,\phi_{ t_1}(t)+c_4) \,dB(t)\label{n2}
\end{equation}

\noindent or equivalently

\begin{equation}
\phi_{ t_1}(t)=\int_0^t\,\left(c_1\,\phi_{ t_1}(s)+c_2\right)\,ds+\int_0^t\,\left(c_3\,\phi_{ t_1}(s)+c_4\right)\,dB(s)\label{n3}
\end{equation}

\noindent where $c_1,c_2,c_3,c_4 $ are constants,  $B(s)$ is one dimensional Brownian motion and the stochastic integral on the right hand side of (\ref{n3}) is in the sense of It\^{o} (cf. [1]). In that case $\phi_{ t_1}(t)$ is actually a stochastic process $\phi_{ t_1}(t,\omega)$ and the problem of finding the optimal switch-off time  $\hat{t_1}$ now amounts to finding the first $t_1$ for which

\[
\int_0^{t_2}\,\mu_{t_1}(s) \,ds   =Q
\]   

\noindent for some finite $t_2\geq t_1$, where $\mu_{t_1}(t)$ denotes the mathematical expectation of 
$\phi_{ t_1}(t,\omega)$ . Since the mathematical expectation of an  It\^{o} stochastic integral with respect to Brownian motion is equal to zero, (\ref{n3}) implies upon taking the expectation of both sides that

\[
\mu_{ t_1}(t)=\int_0^t \,\left(c_1\,\mu_{ t_1}(s)+c_2\right)\,ds 
\]

\noindent which can be solved explicitly and yields the mean energy supply rate

\[
\mu_{ t_1}(t)=\frac{c_2}{c_1}\,\left( e^{c_1\,t}-1 \right)
\]

\noindent where on each interval the constants $c_1$ and $c_2$ are determined by using the initial and tying up conditions. Thus we are reduced to the deterministic problem considered in the examples of Section 2,  but this time for the mean energy supply function. The method extends directly to the case when equations (\ref{n1}) and (\ref{n2}) are replaced by the more general equations

\[
d\phi_{ t_1}(t)=f(t,\phi_{ t_1}(t))\,dt
\]

\noindent and

\[
d\phi_{ t_1}(t)= f(t,\phi_{ t_1}(t)) \,dt+ g(t,\phi_{ t_1}(t))\,dB(t)
\]

\noindent respectively.

\end{document}